\documentstyle[preprint,aps,eqsecnum]{revtex}
\begin{document}
\def\today{\space\number\day\ \ifcase\month\or
January\or February\or March\or April\or May\or June\or July\or
August\or September\or October\or November\or December\fi
\ \number\year}
\overfullrule=0pt  
\def\mynote#1{{[{\it NOTE: #1}]}}
\def\fEQN#1#2{$$\hbox{\it #1\hfil}\EQN{#2}$$}
\def\Acknowledgements{{\bigskip\leftline
{{\bf Acknowledgments}}\medskip}}

\begin{titlepage}

\begin{flushright}
DFTUZ/96-14\\  
hep-th/9605455\\
\end{flushright}

 \vspace{0cm}
 
\begin{center}
{\large\bf  Effective Field Theory of Gravity, Reduction of\\    
Couplings and the Renormalization Group}

 \vspace{0.4cm}
 
{\bf Mario Atance$^{\ast}$}\footnote{E-mail atance@posta.unizar.es}  and
{\bf Jos\'e~Luis~Cort\'es$^\dagger$}\footnote{E-mail cortes@posta.unizar.es}  

 \vspace{0.2cm}

{\sl Departamento de F\'{\i}sica Te\'orica,\\ 
Universidad de Zaragoza,
50009 Zaragoza, Spain.}

 \vspace{0.2cm}

\centerline{May 30, 1996}

\end{center}
\vspace{0.2cm}
\begin{abstract}
The structure of the renormalization group equations for the low
energy effective theory of gravity coupled to a scalar field is
presented. An approximate solution to these equations with a finite
number of independent renormalized parameters can be found 
when the mass scale characteristic of the fluctuations in the geometry
is much smaller than the Planck mass. The cosmological constant problem 
is reformulated in this context and some conditions on the matter field 
content and interactions required in order to have a sufficiently small 
cosmological constant are identified. 
\end{abstract}

\end{titlepage}
\hfill
\section{Introduction}

This work can be classified within the perturbative approach in which
quantum gravity is seen as a theory of small quantum fluctuations 
around a flat Minkowski background spacetime. From this point of view
quantum gravity can be regarded as just another field theory to be
quantised in a standard way as it should be the case for any
relativistic quantum theory at low energy~\cite{Weinberg I}. This 
leads to identify an effective field theory with the gravitational
field described by a symmetric two-index tensor field as the low
energy effective field theoretic formulation of quantum gravity.

Like any effective field theory its Lagrangian density will contain
an infinite number of terms of arbitrary dimensionality and, 
therefore, is not perturbatively renormalizable in the usual power 
counting sense~\cite{Dyson}. But it is renormalizable in the sense 
that all the ultraviolet divergences can be cancelled by a 
renormalization of the infinite number of parameters corresponding 
to the most general action invariant under general covariant 
transformations~\cite{Gomis-Weinberg}. Then the perturbative approach 
of renormalized effective theories~\cite{EFT} can also be applied 
in general relativity.
This point of view has been recently advocated by 
J.F.Donoghue~\cite{Donoghue} who shows how some large distance 
quantum gravitational effects can be derived within this framework.

In order to compute results for physical quantities in an effective 
field theory it is necessary to specify the lagrangian together with a 
renormalization scheme. A natural way to parametrize the lagrangian
is based on the introduction of a fixed mass scale $M$, which is a
characteristic scale of the physical system described by the
effective theory, and a dimensionless parameter for each term in the 
lagrangian giving the corresponding coefficient in units of $M$ raised
to the appropriate power. Then one has an expansion of the lagrangian
with terms of dimensionality greater than four suppressed by 
negative powers of the mass scale $M$. If one wants to have a 
well-defined expansion (with terms of higher dimension being less
important in the calculation of physical quantities) a mass-independent
renormalization scheme must be chosen~\cite{Manohar}.

Each term in the lagrangian of the effective theory of gravity is a 
product of (covariant derivatives of) matter fields and components
of the Riemann tensor. At sufficiently low energies and for 
sufficiently small fluctuations in the matter fields and the geometry 
of space-time, the lower dimensional terms in the lagrangian will be
dominant and the expansion in the effective lagrangian is a good 
approximation. At higher energies, and/or for larger fluctuations of the
matter system or the geometry, higher order terms in the effective 
lagrangian become comparable to the lower dimensional terms. There is 
an intermediate situation where corrections to the dominant term can
be incorporated as a small perturbation. 

In the general case, at each order in the expansion new free parameters 
appear and the predictibility of the effective theory is reduced. But 
this is not always necessarily the case. The renormalization group
equations, fixing the dependence of the renormalized parameters on
the renormalization scale, allow to identify special situations where
only a finite number of renormalized parameters can be chosen freely.
In those cases the predictibility of the theory is not lost when
successive terms in the expansion of the effective theory are 
incorporated.

If one goes beyond the domain of validity of the perturbative approach
to gravity then new interactions and new degrees of freedom will be
required in a new theory beyond quantum field theory based on some
unknown general principles~\cite{Isham}. The main point of this work 
is to investigate the possibility that this new theory going beyond
the perturbative regime is such that its low energy limit is as 
independent on the details of the theory as possible. In other words
we consider an effective field theory with a minimal number of free
renormalized parameters.

In a previous work~\cite{AC} the general structure of the 
renormalization group equations for the effective field theory of
pure gravity was identified. In the limiting case where the mass
scale of the effective theory is much smaller than the Planck mass
(a possibility compatible with the renormalization group
equations), a theory with just one free renormalized parameter is
obtained when contributions suppressed by inverse powers of the
Planck mass are neglected. The aim of this work is to extend these
results to the effective field theory of gravity, including matter
fields and non-gravitational interactions. Additional renormalization
group equations for the new parameters as well as the modifications
induced by these new parameters on the renormalization group equations
of the pure gravity theory have to be considered. Then one has to
identify what are the conditions to be able to express the 
effective lagrangian in terms of a finite number of free parameters
in a way consistent with the renormalization group equations.

In the next section we consider the renormalization of the theory 
of a scalar field coupled to a symmetric two-index tensor field
invariant under general covariant transformations. The
absence of a dimensionfull ultraviolet cut-off in a mass 
independent substraction scheme (dimensional regularization and
minimal substraction) allows us to give the general structure of the
renormalization group equations for the dimensionless renormalized
parameters of the theory. The renormalized parameters corresponding 
to terms in the lagrangian of dimension less than four (mass and 
cosmological term) can be set equal to zero. In this case the 
renormalization group equations for the remaining parameters have a 
triangular structure. For a given term in the lagrangian, the 
renormalization group equation for the 
corresponding parameter depends only on a finite number of
parameters corresponding to the terms in the lagrangian of  
dimensionality smaller or equal than  that of the original term.
The $\beta$ functions are fixed by dimensional 
arguments up to the dependence on the parameter corresponding
to the scalar quartic self-coupling term. There
is a series expansion in this parameter which can be determined
order by order in perturbation theory.

In section 3 a discussion of the possibility to find a renormalized 
lagrangian with a finite number of independent renormalized parameters 
is presented. Two special cases are identified. In the first case 
one has a generalization of the reduction obtained in a previous work
in pure gravity~\cite{AC} with a mass scale $M_{R}$,
characteristic of the fluctuations in the geometry of space-time, 
much smaller than the Planck mass and an additional dependence on the 
scalar self-coupling determined perturbatively. A second case corresponds 
to the presence of still another new independent parameter corresponding 
to a term of dimension six in the matter field lagrangian. This new 
parameter defines a new mass scale, associated to the matter field 
fluctuations, together with the Planck mass and the scale $M_{R}$.
In both cases the reduction of the infinite number of parameters in the
general effective lagrangian in terms of three or four independent
parameters can be determined systematically order by order in
perturbation theory by using the renormalization group equations of
the effective theory.

In section 4 we discuss the modification induced in the 
renormalization group equations when a mass term in the lagrangian
is added. In the general case a cosmological term is unavoidable
and, due to the presence of terms of dimension less than four in the
lagrangian, the renormalization group equations lose its simple structure
and all the infinite parameters appear in the renormalization
group equation of each parameter. The only way to translate to this case 
the discussion of the possibility to have an effective theory with a
finite number of parameters is based on the assumption that the 
dimensionless parameters corresponding to the mass and cosmological
terms in the lagrangian are sufficiently small to treat the modifications 
they induce on the $\beta$ functions of the remaining parameters as a
small perturbation that can be neglected in a first approximation.
In fact the experimental upper bound on the cosmological constant forces 
this to be the case for the cosmological term and the parameter
corresponding to the mass term has to be tuned in such a way that the
scalar mass is much smaller than the remaining mass scales of the 
effective theory. The consistency of the tuning required on the 
cosmological parameter with its renormalization group equation leads 
to a reformulation of the cosmological constant problem~\cite{const.cosmo} 
at the level of the effective theory. Different alternatives to the 
solution of this problem in connection with the reduction of parameters 
in the effective theory of gravity are discussed. 

We end up in section 5 with a summary and some concluding comments.

\section{Renormalization group equations. Massless case}

The main idea in this work is to try to get some information on
quantum gravity from the renormalization group equations of its
effective field theory formulation. This equations can be 
derived following the same steps of a perturbatively renormalizable
theory in the power counting sense~\cite{Gross}. A very important tool 
in the perturbative renormalization of an effective field theory is
dimensional regularization~\cite{DR} and the minimal substraction 
scheme~\cite{t'Hooft}. 

We will consider for definiteness the simplest matter system (a real
scalar field) coupled to the gravitational field but most of the
discussion can be translated directly to a general matter system. It 
is convenient to introduce a fixed mass scale $M$ as a reference unit
for all the couplings of the effective theory. The general expression
for the action can be written, using the invariance under general
covariant transformations, in the following form

\begin{equation} S=\int d^{4}x \, \,
\lbrack {\cal L}_{g} + {\cal L}_{m} + {\cal L}_{n.m.}
\rbrack  \label{action} ,\end{equation}

\noindent where ${\cal L}_{g}$,   is the lagrangian density of the 
gravitational field, and ${\cal L}_{m}$ , ${\cal L}_{n.m.}$ are the 
lagrangian densities for the minimal and  non-minimal coupling of 
the scalar and gravitational fields respectively

\begin{equation} {\cal L}_{g} = \, {\sqrt {-g}} \,
{M^{2} \over \alpha_{1}^{2}}
\lbrack R + 
{{\vec \alpha}_{2} \over M^{2}} \, {\vec R}^{(2)} \, + \,
{{\vec \alpha}_{4} \over M^{4}} \, {\vec R}^{(4)} \, + \, ...
\rbrack  \label{Lg} , \end{equation}

\begin{equation} {\cal L}_{m} = \, {\sqrt {-g}} \,
\lbrack {1\over 2} g^{\mu\nu} \partial_{\mu} \phi
\partial_{\nu} \phi \, + \, \lambda_{0} \, \phi^{4} \, + \,
{{\vec \lambda}_{2} \over M^{2}} \, {\vec {\cal L}}_{m}^{(2)} \, + \,
{{\vec \lambda}_{4} \over M^{4}} \, {\vec {\cal L}}_{m}^{(4)} \, + \, ... 
\rbrack  \label{Lm} , \end{equation}

\begin{equation} {\cal L}_{n.m.} = \, {\sqrt {-g}} \,
\lbrack \xi_{0} R \phi^{2} \, + \,
{{\vec \xi}_{2} \over M^{2}} \, {\vec {\cal L}}_{n.m.}^{(2)} \, + \,
{{\vec \xi}_{4} \over M^{4}} \, {\vec {\cal L}}_{n.m.}^{(4)} \, + \, ... 
\rbrack  \label{Lnm} . \end{equation}

\noindent A cosmological constant (constant term inside the 
brackets in (\ref{Lg}) ) and a mass ($\phi^{2}$ term in (\ref{Lm}) )
have not been included (later we will see how the structure of the
renormalization group equations would be affected in the presence of
such terms) . The coefficients $ \alpha_{1}$, $\vec \alpha_{2}$, 
$\vec \alpha_{4}$,... , $\lambda_{0}$, $\vec \lambda_{2}$,
$\vec \lambda_{4}$,... , $\xi_{0}$, $\vec \xi_{2}$,
$\vec \xi_{4}$,... , are dimensionless parameters.

The dimensionality of the different terms in the action 
fixes the power dependence on the mass scale $M$ of the 
effective theory. In the lagrangian
density of pure gravity ${\cal L}_{g}$ , $ R $ is the
scalar curvature, ${\vec R}^{(2)} $ is a vector with the three different
invariants built out of two Riemann tensors as components, 
the different invariants with three Riemann tensors or two Riemann 
tensors and two derivatives are the components of the vector 
${\vec R}^{(4)}$ and so on. The expansion in inverse powers of the mass 
scale for the terms depending on the matter field has been written
in a compact notation where ${\vec {\cal L}}_{m}^{(2n)}$ is a vector
whose components are the different terms of dimension $4 + 2n$ built 
out of the scalar field and derivatives of the scalar field with all 
derivatives replaced by general covariant derivatives. The additional 
terms of the same dimension involving the Riemann tensor are the 
components of ${\vec {\cal L}}_{n.m.}^{(2n)}$ in the energy expansion
of the nonminimal coupling of the scalar and gravitational fields
(\ref{Lnm}).

For the first terms in the effective field theory expansion one has

\begin{eqnarray}
{\vec R}^{(2)} \,& = &\, \left( 
R^{2} \, , \, R_{\mu \nu} R^{\mu \nu} \, , \, 
\epsilon^{\alpha \beta \gamma \delta}
\epsilon_{\mu \nu \rho \sigma} {R^{\mu \nu}}_{\alpha \beta}
{R^{\rho \sigma}}_{\gamma \delta} \, \right)\nonumber \\
{\vec {\cal L}}_{m}^{(2)} \,& = &\, \left(
( g^{\mu\nu} D_{\mu} \partial_{\nu} \phi ) ^2 \, , \, \phi^{2}
g^{\mu\nu} \partial_{\mu} \phi \partial_{\nu} \phi \, , \,
\phi^{6} \, \right) \label{L2}\\
{\vec {\cal L}}_{n.m.}^{(2)} \,& = &\, \left(
{\vec R}^{(2)} \phi^{2} \, , \, 
R  g^{\mu\nu} \partial_{\mu} \phi \partial_{\nu} \phi \, , \,
R^{\mu\nu} \partial_{\mu} \phi \partial_{\nu} \phi \, , \,
R \phi^{4} \right).\nonumber
\end{eqnarray}

The general parametrization of the effective action is
redundant for two different reasons. First, a change in the
scale $M$ is equivalent to an appropriate rescaling of every
dimensionless parameter. Second, by using a non-linear 
redefinition of fields it is possible to eliminate some of the 
terms in the action (\ref{action})-(\ref{Lnm}). Nevertheless it
is simpler to use this redundant parametrization in order to
identify the general structure of the renormalization group
equations. There are two simple examples for a convenient 
choice of the mass scale of the effective theory $M$. If there
is a choice of the scale $M$ such that all the dimensionless
parameters $\alpha_{2n}^{(i)}$ in the lagrangian density of the 
gravitational field ${\cal L}_{g}$ are simultaneously of order
one then this scale characterizes the size of the fluctuations
in the geometry. Alternatively, if there is a choice of $M$
such that all the dimensionless parameters $\lambda_{2n}^{(i)}$ ,
$\xi_{2n}^{(i)}$ are of order one then this is the scale 
characteristic of the matter field fluctuations. Once the 
scale of the effective theory $M$ has been choosen as one of
the scales of the physical system, using
the Newtonian limit of this action one has 
${\alpha_1}^2 = 16 \pi ( {M \over {M_{Pl}}})^2 $ where $M_{Pl}$ is
the Planck mass. Then $\alpha_{1}$ gives the scale $M$ in units
of the Planck mass.

The simplest case, a system with a unique natural
scale (Planck mass), corresponds to an effective action with
all the dimensionless parameters of order one when $M \sim M_{Pl}$.
Next one can consider a case where all the parameters except 
$\alpha_{1}$ are of order one at a given value of $M$ and it
corresponds to a system with two mass scales, one associated to
the classical limit and a common scale for the geometry
and matter field fluctuations. Another possibility is that there
are two different choices for $M$, one making all $\alpha_{2n}^{(i)}$
of order one and a second one making all $\lambda_{2n}^{(i)}$ ,
$\xi_{2n}^{(i)}$ of order one; in this case there are two different 
scales for the fluctuations of the geometry and the matter system
together with the Planck scale. One could consider even more 
complicated systems with more and more different characteristic 
mass scales corresponding at the level of the effective field
theory to richer hierarchies for the dimensionless parameters.

A perturbative analysis of the action (\ref{action})-(\ref{Lnm})
based on the decomposition of the metric
\begin{equation} g_{\mu \nu} = \eta_{\mu \nu} + {\alpha_1 \over
M} h_{\mu \nu} \,\,\,\,\,\,  \eta_{\mu \nu} = diag(1, -1, -1, -1), 
\end{equation}
can be done~\cite{t'Hooft-Veltman,Goroff-Sagnotti,Veltman} 
by using the standard methods of gauge theories. A non-invariant
term (gauge fixing) has to be added to the action (\ref{action}), 
for example

\begin{equation} S_{g.f.} = \int d^{4}x \,{\sqrt {-g}} \,
{1 \over 2} (\partial^{\rho} h_{\mu\rho} -
{1 \over 2} \partial_{\mu} h^{\rho}_{\rho})^{2}
\label{gf} \end{equation}

\noindent is a very convenient choice for explicit calculations. 
The standard derivation, in perturbatively renormalizable
theories, of the renormalization group in a mass independent
renormalization scheme~\cite{Gross} can be translated to an
effective field theory. One has an infinite number of bare
parameters in one to one correspondence with the dimensionless
parameters of the effective action. Using dimensional 
regularization one has expressions for the bare parameters
in terms of the renormalized parameters with poles when
$ \epsilon \rightarrow 0 $ (dimension $D = 4 - \epsilon $).
From the independence of the bare parameters on the 
renormalization scale $\mu$, one concludes that any change of
$\mu$ must be equivalent to a change in the renormalized
parameters. The renormalization group equations express this
fact. 

In the case of gravity coupled to a scalar field these
equations are

\begin{equation} \mu {d \alpha_{1} \over d \mu} \,=\, 
\beta_{{\alpha}_{1}} ({\vec \alpha} \, , \,{\vec \lambda}
\, , \, \alpha_{1} {\vec \xi} \,) \,\,\, , \,\,\,\,
\mu {d \alpha_{2n}^{(i)} \over d \mu} \,=\,
\beta_{{\alpha}_{2n}^{(i)}} ({\vec \alpha} \, , \,{\vec \lambda}
\, , \, \alpha_{1} {\vec \xi} \,)
\label{RGEg} \,\,\,,\end{equation}

\begin{equation}
\mu {d \lambda_{2n}^{(i)} \over d \mu} \,=\,
\beta_{{\lambda}_{2n}^{(i)}} ({\vec \alpha} \, , \,{\vec \lambda}
\, , \, \alpha_{1} {\vec \xi} \,)
\label{RGEm} \,\,\,,\end{equation}

\begin{equation}
\mu {d {( \alpha_{1} \xi_{2n}^{(i)} )} \over d \mu} \,=\,
\beta_{{\alpha_{1} \xi}_{2n}^{(i)}} ({\vec \alpha} \, , \,{\vec \lambda}
\, , \, \alpha_{1} {\vec \xi} \,)
\label{RGEnm} \,\,\,,\end{equation}

\noindent where ${\vec \alpha}$, ${\vec \lambda}$, ${\vec \xi}$, are 
the set of dimensionless parameters appearing in ${\cal L}_{g}$, 
${\cal L}_{m}$, ${\cal L}_{n.m.}$ respectively. The renormalization 
group $\beta$ functions are determined
perturbatively from the residues of the simple poles at 
$\epsilon = 0$ in the relations between bare and renormalized 
dimensionless parameters. In the parametrization used in 
(\ref{action})-(\ref{Lnm}) any interaction term with the
gravitational field $h_{\mu\nu}$ is proportional to $\alpha_{1}$.
This is the reason why the parameters corresponding to the
non-minimal coupling appear in the renormalization group
equations through the combination $\alpha_{1} {\vec \xi}$.

Dimensional arguments together with the presence of a single
mass scale $M$ ( the dependence on the renormalization scale
$\mu$ is logarithmic ) lead to identify a simple structure
for the renormalization group equations which is a
generalization of the structure found in the case of pure
gravity~\cite{AC}. The $\beta$ functions
satisfy the homogeneity conditions

\begin{equation}
\beta_{{\alpha}_{1}} ({\vec \alpha}' \, , \,{\vec \lambda}'
\, , \, \alpha^{'}_{1} {\vec \xi}' \,) \, = \, 
t \, \beta_{{\alpha}_{1}} ({\vec \alpha} \, , \,{\vec \lambda}
\, , \, \alpha_{1} {\vec \xi} \,) \label{hca1} \,,\end{equation}

\begin{equation}
\beta_{{\alpha}_{2n}^{(i)}} ({\vec \alpha}' \, , \,{\vec \lambda}'
\, , \, \alpha^{'}_{1} {\vec \xi}' \,) \, = \,
t^{2n} \, \beta_{{\alpha}_{2n}^{(i)}} ({\vec \alpha} \, , \,
{\vec \lambda} \, , \, \alpha_{1} {\vec \xi} \,)
\label{hca2n} \,,\end{equation}

\begin{equation}
\beta_{{\lambda}_{2n}^{(i)}} ({\vec \alpha}' \, , \,{\vec \lambda}'
\, , \, \alpha^{'}_{1} {\vec \xi}' \,) \, = \,
t^{2n} \, \beta_{{\lambda}_{2n}^{(i)}} ({\vec \alpha} \, , \,
{\vec \lambda} \, , \, \alpha_{1} {\vec \xi} \,)
\label{hcl2n} \,,\end{equation}

\begin{equation}
\beta_{\alpha_{1} {\xi}_{2n}^{(i)}} ({\vec \alpha}' \, , \,
{\vec \lambda}' \, , \, {\alpha}'_{1} {\vec \xi}' \,) \, = \,
t^{2n+1} \, \beta_{\alpha_{1} {\xi}_{2n}^{(i)}} 
({\vec \alpha} \, , \, {\vec \lambda} \, , \, \alpha_{1} 
{\vec \xi} \,) \label{hcx2n}\,. \end{equation}

\noindent where

\begin{equation}
{\alpha}'_{1} \, = \, t \, \alpha_{1} \,,\,\,\,
{\vec \alpha}'_{2n} \, = \, t^{2n} \, {\vec \alpha}_{2n} 
\, ,  \\
{\vec \lambda}'_{2n} \, = \, t^{2n} \, {\vec \lambda}_{2n} \, , 
\, {\vec \xi}'_{2n} \, = \, t^{2n} \, {\vec \xi}_{2n}
\label{ht}\,. \end{equation}

These conditions put strong restrictions on the dependence 
of the $\beta$ functions on all the dimensionless parameters
with one exception, the scalar self coupling $\lambda_{0}$.
Each renormalization group beta function will be a polinomial
in the remaining parameters of a given degree with 
a series expansion in $\lambda_{0}$ as coefficients
which can be determined order by order in perturbation theory.
From (\ref{hca2n})-(\ref{hcx2n}) one can see that the $\mu$
dependence of ${\vec \alpha}_{2n}$, ${\vec \lambda}_{2n}$,
${\vec \xi}_{2n}$, is fixed by a finite number of parameters
${\vec \alpha}_{k}$, ${\vec \lambda}_{k}$ and
${\vec \xi}_{k}$ with $k \leq 2n$. This triangular structure 
of the renormalization group equations in a mass independent
renormalization scheme is the main ingredient in the discussion 
of the effective field theory formulation of gravity
presented in this work.

A mass for the scalar field and a
cosmological constant term corresponds, in the parametrization
of the effective field theory we are using, to the addition 
of a term $\lambda_{-2} M^{2} \phi^2$ and $\alpha_{-2} M^{2}$
inside the brackets in (\ref{Lm}) and (\ref{Lg}) respectively.
The modifications in the renormalization group equations are
the addition of two equations for $\alpha_{-2}$ and
$\lambda_{-2}$ and the presence of new contributions in
the $\beta$ functions depending on $\alpha_{-2}$, $\lambda_{-2}$.
The homogeneity conditions (\ref{hca1})-(\ref{hcx2n}) can be
generalized including $\alpha^{'}_{-2} \, = \, t^{-2} \,
\alpha_{-2}$ and $\lambda^{'}_{-2} \, = \, t^{-2} \,
\lambda_{-2}$. One consequence of these conditions is that 
$\beta_{{\alpha}_{-2}}$ and $\beta_{{\lambda}_{-2}}$ are both
proportional to $\alpha_{-2}$ or $\lambda_{-2}$ and then a
vanishing mass and cosmological constant parameters, required
in order to get the simple triangular structure for the
$\mu$ dependence of the parameters of the effective action, is
consistent with the renormalization group equations.

To illustrate the general previous considerations we will end up
this section by writing explicitly the renormalization group
equations for the first terms in the ${1 \over M}$ expansion of 
the effective action. One has

\begin{equation}
\mu {d \lambda_{0} \over d \mu} \,=\,
\lambda_{0}^{2} B_{0}^{(1)} \label{RGEl0} \,,\end{equation}

\begin{equation}
\mu {d \xi_{0} \over d \mu} \,=\,
\lambda_{0}  B_{0}^{(2)}  (\, 1 \, - \,{{\xi_{0}} \over 12} \, )
\label{RGEx0} \,,\end{equation}

\noindent where $B_{0}^{(k)}$ are power expansions in $\lambda_{0}$.
There is no effect of the gravitational interaction on the
$\mu$ dependence of the scalar self-coupling and 
(\ref{RGEl0}) is the renormalization group equation of the
$\lambda \phi^{4}$ theory. The $\beta$ function for $\xi_{0}$
can be read from the insertion of an external graviton and 
a single gravitational coupling on the scalar self-energy and
it does not involve any effect from the fluctuations of the
gravitational field. This is the reason why there is a
common power expansion in $\lambda_{0}$ for the two terms in
the $\beta$ function for $\xi_{0}$, and the relative coefficient,
$-12$, can be read from the one loop results for the 
divergencies of gravitation interacting with a scalar 
particle~\cite{t'Hooft-Veltman}.

Next one has

\begin{equation} \mu {d \alpha_{1} \over d \mu} \,=\, 0
\label{RGEa1}.\end{equation}

\noindent This is because any gravitational interaction
is proportional to $\alpha_{1}$ and it is not posssible to 
have a divergent ${1 \over \epsilon}$ contribution 
compatible with (\ref{hca1}). At order ${1 \over M^{2}}$ the
renormalization group equations are

\begin{equation}
\mu {d \alpha_{2}^{(i)} \over d \mu} \,=\, \alpha_{1}^{2} \,
\lbrack B_{2}^{(i,1)} \,+\, \xi_{0} B_{2}^{(i,2)} \,+\,
\xi_{0}^{2} B_{2}^{(i,3)} \rbrack \label{RGEa2} \,,\end{equation}

\begin{equation}
\mu {d \lambda_{2}^{(i)} \over d \mu} \,=\, \sum_{j} 
\lambda_{2}^{(j)} \lambda_{0} B_{2}^{(i,j,4)} \, + \,
{\alpha_{1}}^{2} \lambda_{0} \lbrack B_{2}^{(i,5)} \,+\, 
\xi_{0} B_{2}^{(i,6)} \,+\, {\xi_{0}}^{2} B_{2}^{(i,7)} \rbrack
\label{RGEl2} \,,\end{equation}

\begin{eqnarray}
\mu {d \xi_{2}^{(i)} \over d \mu} \,&=&\, \sum_{j} 
\xi_{2}^{(j)} \lambda_{0} B_{2}^{(i,j,8)} \, + \,
\sum_{j} \lambda_{2}^{j} \lbrack B_{2}^{(j,9)} \,+\,
\xi_{0} B_{2}^{(j,10)} \rbrack \, + \,\nonumber\\
&& {\alpha_{1}}^{2} \, \lbrack \,
B_{2}^{(i,11)} \,+\, \xi_{0} B_{2}^{(i,12)} \,+\, 
{\xi_{0}}^{2} B_{2}^{(i,13)} \,+\, {\xi_{0}}^{3} B_{2}^{(i,14)} \,
\rbrack \label{RGEx2} \,,\end{eqnarray}

\noindent where once more one has several coefficients
$B_{2}$ as power expansions in the self-coupling $\lambda_{0}$.
The generalization to higher orders terms in the ${1 \over M}$
expansion is obvious.

The zero order term in the expansion of $B_{2}^{(i,1)}$,
$B_{2}^{(i,2)}$, $B_{2}^{(i,3)}$ for the gauge fixing 
in (\ref{gf}) can be read from the one loop
calculations in~\cite{t'Hooft-Veltman,Goroff-Sagnotti}.
The determination of the renormalization group coefficients for 
$\alpha_{4}^{(i)}$ in lowest order requires a two loop
calculation whose results in the case of pure gravity
are given in~\cite{Goroff-Sagnotti}. Although the divergences, 
and then the coefficients of the renormalization group equations,  
depend on the gauge fixing condition~\cite{rusos}, there are no
additional substractions and a modification of the gauge
fixing condition simply gives an equivalent formulation of
the effective field theory.

The general structure of the renormalization group equations
based on dimensional arguments will be valid beyond the 
simple example of a scalar field considered in this section.
The only difference will be on the number and explicit form of
the different terms in the minimal and non-minimal coupling
of the matter system to the gravitational field and on the 
values for the coefficients of the counterterms which determine 
the coefficients in the renormalization group equations.

\section{Reduction of couplings}

\subsection{General considerations}

The idea of looking for relations between the couplings of 
a renormalizable field theory which are independent of the 
renormalization scale and compatible with the renormalizability
of the theory has been studied in recent years for different 
purposes~\cite{Oehme}. The program of reduction of couplings was
initiated in~\cite{Oehme-Zimmerman} by looking for massless
renormalizable theories in the power counting sense with a
single dimensionless coupling parameter. The same basic ideas
were considered in~\cite{Perry-Wilson} in order to understand the 
presence of a finite number of parameters in a renormalizable theory
despite the appearence of an infinite number of interaction terms
in the light-cone quantization method. The underlying Lorentz
covariance and gauge invariance, which are not manifest in this
quantization scheme, are reflected in the possibility to determine
the renormalization group trajectories in terms of a finite
number of running variables.

In the case of the effective field theory formulation of gravity
one also has an infinite number of interaction terms as required
by the ultraviolet divergences of the theory but one can consider
the possibility to have a finite number of independent renormalized 
parameters. The reduction of couplings in this case could be a 
consequence of a symmetry of the underlying fundamental theory 
which is hidden in the field theoretical limit. If the effective 
field theory of gravity is a result of the application of the 
reduction program then one has a situation where the field 
theoretical approach goes as far as possible in the sense that 
the low energy limit of the theory is only sensitive to the 
details of the underlying theory through the value of a finite 
number of parameters.
 
This idea has been studied in the simpler case of pure gravity 
in~\cite{AC}. The result is that it is possible to
express all the parameters ${\vec \alpha}_{2n}$ in terms of two
parameters $\alpha_{1}$ and $\alpha_{2}$ in a way compatible
with the renormalization group equations. But in the general case
there are as many free parameters in these relations as parameters
${\vec \alpha}_{2n}$ due to the trivial $\mu$ dependence of 
$\alpha_{1}$. Then one has a real reduction of couplings only when
one assumes that $\alpha_{1}^2 \ll \alpha_{2}$, which corresponds to 
a mass scale for the fluctuations of the geometry much smaller than 
the Planck mass, and keeps only the dominant term in the expansion
in powers of $\alpha_{1}$. The aim of this section is to consider
possible extensions of the reduction of couplings in the
presence of a scalar matter field coupled to the gravitational
field. The main difference with the pure gravity case is the 
presence of a parameter, $\lambda_{0}$, in the renormalization
group equations and in the relations between couplings. The
dependence on this new parameter is not fixed by dimensional 
considerations and is incorporated order by order in a 
perturbative expansion. Another difference comes from the
possibility to consider new reductions of couplings with an
additional independent parameter $\lambda_{2}$ associated
to the matter system.

\subsection{ Minimal reduction}

The triangular structure of the renormalization group 
equations allows to look for reduction of parameters order
by order in the ${1 \over M}$ expansion of the effective
theory. In lowest order the relevant parameters to consider 
are $\lambda_{0}$ , $\xi_{0}$ and the renormalization
group equations (\ref{RGEl0}) , (\ref{RGEx0}) . A reduction 
at this level corresponds to express $\xi_{0}$ as a power
expansion in $\lambda_{0}$ with coefficients fixed in order
to reproduce its $\mu$ dependence. The only posssible 
reduction is a trivial one, $\xi_{0} \, = \, {1 \over 12}$.

At order ${1\over M^{2}}$ the first parameter to consider is
${\vec \alpha}_{2}$. The renormalization group equation
(\ref{RGEa2}) can be written, using the reduction of 
$\xi_{0}$, as

\begin{equation}
\mu {d {\vec \alpha}_{2} \over d \mu} \,=\, 
{\vec A}_{2} \, \alpha_{1}^{2}
\label{RGE2a2} \,.\end{equation}

\noindent The factorization at this level of the radiative 
corrections due to the matter field leads to identify a unique 
power expansion in $\lambda_{0}$, $A_{2}$, 

\begin{equation}
{\vec A}_{2} \, = \, {\vec a}_{2}^{(0)} \, A_{2}
\label{fact.} \,,\end{equation}

\noindent where the zero order term in $A_{2}$ is chosen to
be one and the constant coefficients ${\vec a}_{2}^{(0)}$ can be 
read from the one loop result~\cite{t'Hooft-Veltman} for
the counterterms quadratic in the Riemann tensor. A reduction
of couplings at this level can be obtained if one introduces
a dimensionless parameter $\alpha_{2}$ with a renormalization
scale dependence given by 

\begin{equation}
\mu {d \alpha_{2} \over d \mu} \,=\, A_{2} \, \alpha_{1}^{2}
\label{RRGEa2} \, . \end{equation}

\noindent Then it is possible to write the parameters 
$\alpha_{2}^{(i)}$ in terms of $\alpha_{2}$ and $\alpha_{1}$ in 
a way compatible with the renormalization group equations,

\begin{equation}
{\vec \alpha}_{2} \, = \, {\vec a}_{2}^{(0)} \, \alpha_{2} \, +
{\vec a}_{2}^{(1)} \alpha_{1}^{2}
\label{REa2} \, , \end{equation}

\noindent but, as a consequence of the independence of $\alpha_{1}$
on the renormalization scale, the coefficients ${\vec a}_{2}^{(1)}$
are free parameters and the relations (\ref{REa2}) are not real
reduction equations but simply a reparametrization. If
$\alpha_{1}^{2} \, \ll \, \alpha_{2}$, and one takes the dominant
term in the expansion in powers of ${\alpha_{1}^{2} \over \alpha_{2}}$
in (\ref{REa2}) , then one has a reduction of couplings for the 
terms quadratic in the Riemann tensor.

The next step is to consider the renormalization group equations for
the parameters corresponding to terms of dimension 6 in the minimal
coupling of the scalar and gravitational fields. Using the reduction
of ${\vec \alpha}_{2}$ and $\xi_{0}$ one has

\begin{equation}
\mu {d \lambda_{2}^{(i)} \over d \mu} \,=\,
\sum_{j=1}^3 \lambda_{2}^{(j)} \, \lambda_{0} L_{2}^{(i,j)} \, + \,
L_{2}^{(i)} \, \alpha_{1}^{2}
\label{RRGEl2} \,.\end{equation}

\noindent If one does not consider additional free parameters
(minimal reduction) then one has to be able to express 
$\lambda_{2}^{(i)}$ in terms of $\lambda_{0}$, $\alpha_{1}$
and $\alpha_{2}$ . Consistency with the renormalization group 
equations leads to 

\begin{equation}
{\vec \lambda}_{2} \, = \, {{\vec l}_{2} \over \lambda_{0}} \,
\alpha_{1}^{2} \label{REl2} \,,\end{equation}

\noindent where ${\vec l}_{2}$ are power expansions in
$\lambda_{0}$ with coefficients determined order by order as
the solution of consistency equations which reduce to a
linear system of equations. 

If one inserts the reduction of parameters in the previous
steps into the renormalization group equations (\ref{RGEx2}) 
for the nonminimal couplings of dimension 6 , one obtains

\begin{equation}
\mu {d \xi_{2}^{(i)} \over d \mu} \,=\,
\sum_{j=1}^3 \xi_{2}^{(j)} \, \lambda_{0} X_{2}^{(i,j)} \, + \,
{X_{2}^{(i)} \over \lambda_{0}} \, \alpha_{1}^{2}
\label{RRGEx2} \,.\end{equation}

\noindent Once more the reduction of the parameters $\xi_{2}^{(i)}$
is uniquely determined by the renormalization group equations,

\begin{equation}
{\vec \xi}_{2} \, = \, {{\vec x}_{2} \over \lambda_{0}^{2}} \,
\alpha_{1}^{2} \label{REx2} \,,\end{equation}

\noindent where each term in the expansion in powers of $\lambda_{0}$
of ${\vec x}_{2}$ is the solution of a linear system of equations.

The iterative procedure to reduce parameters can be applied to the
remaining parameters of the effective theory. For the terms of
order ${1 \over M^4}$ in ${\cal L}_{g}$ one has 

\begin{equation}
\mu {d {\vec \alpha}_{4} \over d \mu} \,=\,
{\vec A}_{4}^{(0)} \alpha_{1}^{2} \alpha_{2} \, + \,
{{\vec A}_{4}^{(1)} \over \lambda_{0}^2} \alpha_{1}^{4} 
\label{RRGEa4} \,,\end{equation}

\noindent and a reduction consistent with the renormalization 
group leads to

\begin{equation}
{\vec \alpha}_{4} \, = \, {\vec a}_{4}^{(0)} \alpha_{2}^{2} \, +
{{\vec a}_{4}^{(1)} \over \lambda_{0}} \alpha_{1}^{2} \alpha_{2} \, +
{{\vec a}_{4}^{(2)} \over \lambda_{0}^3} \alpha_{1}^{4}
\label{REa4} \,.\end{equation}

\noindent The coefficients ${\vec a}_{4}^{(k)}$ , $k \,= \, 0,1,2$ ,
are power expansions in $\lambda_{0}$ determined by a system of
linear equations with one exception, the coefficient of
$\lambda_{0}^3$ in ${\vec a}_{4}^{(2)}$, which does not appear in
the consistency equations. Then once more the relations
(\ref{REa4}) are not a real reduction but a reparametrization in
terms of the arbitrary constant coefficients of $\alpha_{1}^{4}$.

For the ${1 \over M^4}$ terms in ${\cal L}_{m}$ one has the 
reduced renormalization group equations

\begin{equation}
\mu {d \lambda_{4}^{(i)} \over d \mu} \,=\,
\sum_{j} \lambda_{4}^{(j)} \, \lambda_{0} L_{4}^{(i,j)} \, + \,
L_{4}^{(i)} \, \alpha_{1}^{2} \alpha_{2} \, + \, 
{{L^{'}}_{4}^{(i)} \over \lambda_{0}^{2}} \, \alpha_{1}^{4}
\label{RRGEl4} \,,\end{equation}

\noindent and the reduction relations

\begin{equation}
{\vec \lambda}_{4} \, = \, 
{{\vec l}_{4}^{(0)} \over \lambda_{0}} \alpha_{1}^{2} \alpha_{2} \, +
{{\vec l}_{4}^{(1)} \over \lambda_{0}^3} \alpha_{1}^{4}
\label{REl4} \,,\end{equation}

which can be used to extend the reduction to the non-minimal
couplings

\begin{equation}
{\vec \xi}_{4} \, = \, 
{{\vec x}_{4}^{(0)} \over \lambda_{0}^2} \alpha_{1}^{2} \alpha_{2} \, +
{{\vec x}_{4}^{(1)} \over \lambda_{0}^4} \alpha_{1}^{4}
\label{REx4} \,.\end{equation}

The steps followed in the determination of the reduction of the
parameters corresponding to terms of order ${1 \over M^4}$ can
be repeated order by order in the expansion in ${1 \over M}$ to get
the minimal reduction of the field theory of gravity coupled to a
scalar field which is given by the relations

\begin{eqnarray}
{\vec \alpha}_{2n} \, &=& \, {\vec a}_{2n} \, \alpha_{2}^{n} \,
\lbrack \, 1 \, + \, {\cal O} ({\alpha_{1}^{2} \over \alpha_{2}})
\rbrack \nonumber\\
{\vec \lambda}_{2n} \, &=& \, {{\vec l}_{2n} \over \lambda_{0}}
\, \alpha_{2}^{n-1} \, \alpha_{1}^{2}  
\lbrack \, 1 \, + \, {\cal O} ({\alpha_{1}^{2} \over \alpha_{2}})
\rbrack \label{MR} \, \\
{\vec \xi}_{2n} \, &=& \, {{\vec x}_{2n} \over \lambda_{0}^{2}}
\, \alpha_{2}^{n-1} \, \alpha_{1}^{2}  
\lbrack \, 1 \, + \, {\cal O} ({\alpha_{1}^{2} \over \alpha_{2}})
\rbrack \nonumber .\end{eqnarray}

\noindent The condition to have a real reduction of couplings
is that $\alpha_{1}^{2} \ll \alpha_{2}$. In this case we have 
shown that there is a unique way to find perturbatively a
unique lagrangian with the property that all the couplings
can be written, at any renormalization scale $\mu$, in terms of 
three independent dimensionless parameters $\lambda_{0}$, 
$\alpha_{1}$ and $\alpha_{2}$. The reduction equations (\ref{MR})
as well as the $\mu$-dependence of the independent parameters
(\ref{RGEl0}), (\ref{RGEa1}), (\ref{RRGEa2}) can be determined
from a perturbative calculation of counterterms.

Each of the parameters of the reduced theory allows to identify 
a typical scale of the system. The scale associated to the 
parameter $\lambda_{0}$ can be taken for example as the scale
$M_{0}$ such that 
${{\lambda_{0} (\mu = M_{0})} \over 4\pi} \, = \, 1$ and it
fixes the range of validity of the perturbative approach to
the effective theory. If we choose the scale $M$ in the energy
expansion of the effective theory as $M_{0}$ then the second
parameter $\alpha_{1}$ is associated to the scale characteristic 
of the classical limit, the Planck mass $M_{Pl}$ through the 
relation ${\alpha_{1}^2 \over 16\pi} \, = \, 
{M_{0}^2 \over M_{Pl}^2}$. The parameter $\alpha_{2} (M_{0})$ 
allows to introduce a third scale in the system which is the 
scale characteristic of the fluctuations of the geometry $M_{R}$
generated by the matter system at the scale $M_{0}$. From the 
expansion of ${\cal L}_{g}$ in powers of the Riemann tensor 
one gets $\alpha_{2} (M_{0}) \, = \,{M_{0}^2 \over M_{R}^2}$  and
the condition for the validity of the reduction 
$\alpha_{1}^{2} \ll \alpha_{2}$ in terms of the three scales of 
the system corresponds to the condition that $M_{R} \ll M_{Pl}$.

\subsection{ Non-minimal reduction}

The minimal reduction of couplings is an extension of the reduction
found in the case of pure gravity in~\cite{AC} with the 
addition of the self-coupling $\lambda_{0}$ to the independent
parameters $\alpha_{1}$ and $\alpha_{2}$. An alternative to this 
reduction corresponds to introduce still another independent
parameter from the matter system lagrangian. This can be done by
going back to the renormalization group equations for
$\lambda_{2}^{(i)}$ (\ref{RRGEl2}) and introducing a new independent
parameter $\lambda_{2}$ with a renormalization scale dependence
given by 

\begin{equation}
\mu {d \lambda_{2} \over d \mu} \,=\, L_{2} \lambda_{0} \, \lambda_{2}
\label{RGERl2} \, , \end{equation}

\noindent where $L_{2}$ is a new expansion in powers of $\lambda_{0}$
to be conveniently choosen in a way compatible with the renormalization
group equations. The most general form for the reduction of the 
parameters $\lambda_{2}^{(i)}$ compatible with dimensional arguments
has three terms proportional to $\lambda_{2}$, $\alpha_{2}$ and 
$\alpha_{1}^{2}$ respectively with coefficients, depending on 
$\lambda_{0}$. These coefficients can be determined perturbatively 
through the consistency conditions on the reduction imposed by the
renormalization group equations. The solution can be written in
the form

\begin{equation}
\lambda_{2}^{(i)} \, = \, {l_{2,0}^{(i)}} \, \lambda_{2} \, + \,
{{l_{2,1}^{(i)}} \over \lambda_{0}} \, \alpha_{1}^{2}
\label{RE2l2} \,. \end{equation}

\noindent The parameter $\lambda_{2}$ can be chosen such that 
$L_{2}$ is a constant corresponding to one of the eigenvalues of
the matrix given by the zero order term in the expansion of
$L_{2}^{(i,j)}$ appearing in the renormalization group equation
of $\lambda_{2}^{(i)}$ (\ref{RRGEl2}). The zero order terms in the
expansion in powers of $\lambda_{0}$ of ${l_{2,0}^{(i)}}$ are 
the components of the corresponding eigenvector. The remaining 
terms in the expansion of the coefficients in the reduction,
${l_{2,0}^{(i)}}$ and ${l_{2,1}^{(i)}}$, are the solutions
of the remaining consistency conditions which reduce to a system of 
linear equations.

Once the new independent parameter $\lambda_{2}$ has been introduced,
with its corresponding renormalization group equation (\ref{RGERl2}),
there is no difficulty to repeat the step by step reduction of
the remaining dimensionless parameters in the effective theory. The
results for the first terms in the ${1 \over M}$ expansion are

\begin{eqnarray}
{\vec \xi}_{2} \, &=& \, {{\vec x}_{2,0} \over \lambda_{0}} \, 
\lambda_{2} \, + \, {{\vec x}_{2,1} \over \lambda_{0}^2} \,
\alpha_{1}^{2} \nonumber \,,\\
{\vec \alpha}_{4} \, &=& \, {\vec a}_{4,0} \, \alpha_{2}^{2} \,+\,
{{\vec a}_{4,1} \over \lambda_{0}} \, \alpha_{2} \alpha_{1}^{2} \,+\,
{{\vec a}_{4,2} \over \lambda_{0}^2} \, \lambda_{2} \alpha_{1}^{2} \,+\,
{{\vec a}_{4,3} \over \lambda_{0}^3} \, \alpha_{1}^{4} \nonumber \,,\\
{\vec \lambda}_{4} \, &=& \,{{\vec l}_{4,0} \over \lambda_{0}}\,
\lambda_{2}^{2} \,+\, {{\vec l}_{4,1} \over \lambda_{0}^2}\,
\lambda_{2} \alpha_{1}^{2} \,+\, 
{{\vec l}_{4,2} \over \lambda_{0}} \alpha_{2} \alpha_{1}^{2} \,+\,
{{\vec l}_{4,3} \over \lambda_{0}^3} \, \alpha_{1}^{4}
\label{NMR} \,,\\
{\vec \xi}_{4} \, &=& \,{{\vec x}_{4,0} \over \lambda_{0}^2}\,
\lambda_{2}^{2} \,+\, {{\vec x}_{4,1} \over \lambda_{0}^3}\,
\lambda_{2} \alpha_{1}^{2} \,+\, 
{{\vec x}_{4,2} \over \lambda_{0}^2} \alpha_{2} \alpha_{1}^{2} \,+\,
{{\vec x}_{4,3} \over \lambda_{0}^4} \, \alpha_{1}^{4}
\nonumber \,.\end{eqnarray}

\noindent The expansion in powers of $\lambda_{0}$ of the
reduction coefficients ${\vec a}$, ${\vec l}$,${\vec x}$, is 
fixed by a set of linear systems of equations with the exception 
of the term proportional to $\lambda_{0}^3$ in ${\vec a}_{4,3}$ 
which remains undetermined. Once more a real reduction of couplings 
requires to assume that $\alpha_{1}^{2} \ll \alpha_{2}$ and to 
neglect terms supressed by powers of ${\alpha_{1}^{2} \over \alpha_{2}}$ 
in the reduction equations (\ref{NMR}).

The new independent parameter $\lambda_{2}$ in this nonminimal
reduction corresponds to a new mass scale $M_{E}$ in the energy 
expansion of the matter system related to the dimensionless 
parameter $\lambda_{2}$ by $\lambda_{2} (M_{0}) \, = \, 
{M_{0}^2 \over M_{E}^2}$. One can see the nonminimal reduction
as an extension of the scalar effective field theory with two
free parameters $\lambda_{0}$ and $\lambda_{2}$. The gravitational
interaction introduces at least two additional parameters 
$\alpha_{1}$ and $\alpha_{2}$. The main difference with respect to
the minimal reduction discussed in the previous subsection is that 
the contributions of higher dimensional operators in the matter
system lagrangian are not necessarily suppressed by inverse 
powers of the Planck mass.

\section{Massive case. Cosmological constant problem}

If one considers the gravitational interaction of a massive
spinless particle then one has to include a term 
$\lambda_{-2} M^{2} \phi^{2}$ in the matter lagrangian density.
The dimensionless parameter $\lambda_{-2}$ has to be taken into
account in the discussion based on dimensional arguments leading 
to the general structure of the renormalization group equations.
The homogeneity conditions of the $\beta$ functions for the 
parameters $\alpha_{1}$, ${\vec \alpha}_{2n}$, ${\vec \lambda}_{2n}$,
${\vec \xi}_{2n}$ include the rescaling of the additional parameter,
$\lambda^{'}_{-2} \, = \, t^{-2} \, \lambda_{-2}$, and the simple
triangular structure is lost due to the contributions proportional
to positive powers of $\lambda_{-2}$ which will be accompanied by 
the parameters corresponding to terms of higher dimensionality. If 
one wants the reduction of couplings to be applicable also in this 
case then one has to assume that the dimensionless parameter
$\lambda_{-2}$ is sufficiently small to treat its effects as a
small perturbation. The additional parameter is associated to a new
mass scale in the system, the scalar particle mass 
$m^2  = \lambda_{-2} M^{2}$, and the reduction of couplings 
requires this scale to be much smaller than any of the other mass
scales of the system. The reduction of couplings identified in
the previous section for the massless case can be taken as the 
zero order term of an expansion of the reduction equations
in powers of the parameter $\lambda_{-2}$.

Another effect of the mass of the scalar particle is that one has
to consider simultaneously a cosmological constant. This can be
seen from the renormalization group equation for the dimensionless 
parameter $\alpha_{-2}$, corresponding to the cosmological term,
which can be written in the form

\begin{equation}
\mu {d \alpha_{-2} \over d \mu} \,=\, {\lambda_{-2}}^2 
\alpha_{1}^2 {\hat \beta}_{{\alpha}_{-2}} (\lambda_{-2},{\vec \alpha},
{\vec \lambda},{\vec \xi})  \, + \, {\cal O} (\alpha_{-2})
\label{RGEa-2} \,,\end{equation}

\noindent Dimensional arguments and the fact that all the 
couplings of the scalar field to the gravitational field
$h_{\mu\nu}$ are proportional to $\alpha_{1}$ have
been used to write the contribution to the $\beta$ function of
the dimensionless cosmological parameter due to the remaining 
parameters in terms of the function ${\hat \beta}_{{\alpha}_{-2}}$ 
which satisfies the homogeneity condition

\begin{equation}
{\hat \beta}_{{\alpha}_{-2}} ({\vec \alpha}' \, , \,
{\vec \lambda}' \, , \, {\vec \xi}' \,) 
\, = \, {\hat \beta}_{{\alpha}_{-2}} \,  ({\vec \alpha} 
\, , \, {\vec \lambda} \, , \, {\vec \xi} \,)
\label{hca-2} \,.\end{equation}

The first terms in the expansion in powers of $\lambda_{-2}$ for
the renormalization group equation of the dimensionless 
cosmological parameter are

\begin{eqnarray}
\mu {d \alpha_{-2} \over d \mu} &=& {\lambda_{-2}}^2 
\alpha_{1}^2 \lbrace B_{-2}^{(1)} \, + \, \lbrack 
\sum_{i} B_{-2}^{(i,1)} \lambda_{2}^{(i)} \,+\,
\sum_{i} B_{-2}^{(i,2)} \xi_{2}^{(i)} \,+\,
B_{-2}^{(2)} \alpha_{1}^2 \rbrack \lambda_{-2} \,+\,\nonumber\\
&&\lbrack \sum_{i} B_{-2}^{(i,3)} \lambda_{4}^{(i)} \,+\,
\sum_{i} B_{-2}^{(i,4)} \xi_{4}^{(i)} \,+\,
\sum_{i,j} B_{-2}^{(i,j,1)} \lambda_{2}^{(i)} \lambda_{2}^{(j)} \,+\,
\nonumber\\
&&\sum_{i,j} B_{-2}^{(i,j,2)} \lambda_{2}^{(i)} \xi_{2}^{(j)} \,+\,
\sum_{i,j} B_{-2}^{(i,j,3)} \xi_{2}^{(i)} \xi_{2}^{(j)} \,+\,
\sum_{i} B_{-2}^{(i,5)} \alpha_{2}^{(i)} \alpha_{1}^{2} \,+\,
\nonumber\\
&&\sum_{i} B_{-2}^{(i,6)} \lambda_{2}^{(i)} \alpha_{1}^{2} \,+\,
\sum_{i} B_{-2}^{(i,7)} \xi_{2}^{(i)} \alpha_{1}^{2} \,+\,
B_{-2}^{(3)} \alpha_{1}^4 \rbrack \lambda_{-2}^2 \,+\
\nonumber\\
&&{\cal O} (\lambda_{-2}^3) \rbrace \, +\, {\cal O} (\alpha_{-2})
\label{RGE2a-2} \,.\end{eqnarray}

The cosmological constant parameter 
$\Lambda \,=\,{\alpha_{-2} \over \alpha_{1}^2} M^4$ can be written 
in units of the scalar mass by using the dimensionless parameters
$\alpha_{1}$, $\alpha_{-2}$ and $\lambda_{-2}$ as
\begin{equation}
{\Lambda \over m^4} \,=\, {\alpha_{-2} \over 
{\alpha_{1}^2 \lambda_{-2}^2}} \,=\, {\alpha_{-2} \over \lambda_{-2}}
{1 \over {\alpha_{1}^2 \lambda_{-2}}} \nonumber \,.\end{equation}

\noindent This, together with the renormalization group equation
(\ref{RGE2a-2}), gives a reformulation of the cosmological constant 
problem at the level of the effective field theoretic formulation.
It is not possible to have $\alpha_{-2} \ll
\alpha_{1}^{2} {\lambda_{-2}}^{2}$ over the energy range of
validity of the effective theory and then it is not possible to have
${\Lambda \over m^4} \ll 1$ as required by the experimental upper
bound for the cosmological constant. 

In order to discuss possible solutions to the cosmological constant 
problem one has first to consider, instead of a scalar field, a
realistic matter system including fermion and gauge fields with 
the strong and electroweak interactions of the minimal standard
model of particle physics. If the matter system content of fields
and interactions is such that as a consequence of a symmetry 
${\hat \beta}_{{\alpha}_{-2}} \, = \, 0$, then

\begin{equation} 
\mu {d \alpha_{-2} \over d \mu} \propto \alpha_{-2} 
\nonumber \,,\end{equation}

\noindent and the cosmological parameter can be made arbitrarily
small\footnote{ A supersymmetric extension of the minimal standard
model is a candidate but it is not clear how the required soft 
supersymmetry breaking terms can be made compatible with the 
vanishing of ${\hat \beta}_{{\alpha}_{-2}}$.}.

Taking into account that $\alpha_{1}^{2} \lambda_{-2} \,=\, {m^2 \over 
M_{Pl}^2}$ is of the order of $10^{-32}$ for a mass scale of the matter
system $m$ of the order of 1 Tev, another possibility to solve the
cosmological problem is based on the assumption that the first two 
terms in the expansion of $\beta_{\alpha_{-2}}$ vanish. The remaining 
terms can be made compatible with the smallness of the cosmological
constant if the combinations of parameters, appearing in the 
coefficient of ${\lambda_{-2}}^2$ inside the brackets in (\ref{RGEa-2}),
are not much bigger than $\alpha_{1}^4$.

If one considers a minimal reduction of the effective theory for a 
realistic matter system with the same number of independent 
parameters $\lambda_{-2}$, $\lambda_{0}$, $\alpha_{1}$, $\alpha_{2}$
as in the case of a scalar field, then the renormalization group
equation of the cosmological parameter will be given by

\begin{eqnarray}
\mu {d \alpha_{-2} \over d \mu} \,=\, {\lambda_{-2}}^2 
\alpha_{1}^2 &&\lbrace A_{-2}^{(1)} \, + \, 
A_{-2}^{(2)} \alpha_{1}^{2} \lambda_{-2} \,+\,
A_{-2}^{(3)} \alpha_{1}^{2} \alpha_{2} \lambda_{-2}^2 \,+\nonumber\\
&& A_{-2}^{(4)} \alpha_{1}^{4} {\lambda_{-2}}^2 \,+\
{\cal O} (\lambda_{-2}^3) \rbrace \, +\, {\cal O} (\alpha_{-2})
\label{MRRGEa-2} \,,\end{eqnarray}

\noindent where the coefficients $A_{-2}^{(k)}$ are power expansions
in the parameter $\lambda_{0}$. In this case the possible solution of 
the cosmological constant problem based on the vanishing of the first
few terms in the $\lambda_{-2}$ expansion of the $\beta$-function
gives just two conditions on the matter system,

\begin{equation}
A_{-2}^{(1)} \, = \, 0 \,\,\,\, A_{-2}^{(2)} \, = \, 0 
\label{MRSCCP} \,. \end{equation}

\noindent Assuming that the remaining coefficient expansions 
are of order one, one has the additional condition for the 
independent dimensionless parameters ${\alpha_{2} \over \alpha_{1}^{2}}
\leq 10^6$ which corresponds to the restriction on the mass scales 
of the effective theory ${M_{R}^2 \over M_{Pl}^2} \geq 10^{-6}$ . On 
the contrary, if the solution to the cosmological problem is based 
on some symmetry of the underlying theory leading to 
$A_{-2}^{(k)} \,=\, 0$ there is no restriction, due to the 
smallness of the cosmological constant, on the mass scales of the 
effective theory.

The cosmological constant problem can be discussed along the same 
lines if there is a nonminimal reduction of the effective theory 
with an additional parameter $\lambda_{2}$ . In this case one
has the renormalization group equation for the dimensionless
cosmological parameter

\begin{eqnarray}
\mu {d \alpha_{-2} \over d \mu} \,&=&\, {\lambda_{-2}}^2 
\alpha_{1}^2 \lbrace {A^{'}_{-2}}^{(1)} \, + \, 
\lbrack {A^{'}_{-2}}^{(2)} \alpha_{1}^{2} \,+\,
{A^{'}_{-2}}^{(3)} \lambda_{2} \rbrack \lambda_{-2} \,+\,\nonumber\\
&& \,+\ \lbrack {A^{'}_{-2}}^{(4)} \lambda_{2}^2 \,+\,
{A^{'}_{-2}}^{(5)} \alpha_{2} \alpha_{1}^{2} \,+\,
{A^{'}_{-2}}^{(6)} \lambda_{2} \alpha_{1}^{2} \,+\,
{A^{'}_{-2}}^{(7)} \alpha_{1}^{4} \rbrack {\lambda_{-2}}^2 \,+\
\nonumber\\
&& \,+\
{\cal O} (\lambda_{-2}^3) \rbrace \, +\, {\cal O} (\alpha_{-2})
\label{NMRRGEa-2} \,,\end{eqnarray}

\noindent In this case a solution to the cosmological constant
problem based on the vanishing of the first two terms in the 
expansion in powers of $\lambda_{-2}$ requires that the matter 
system satisfies the three conditions

\begin{equation}
{A^{'}_{-2}}^{(k)} \, = \, 0 \,\,\, k \, = \, 1, 2, 3.
\label{NMRSCCP} \end{equation}

\noindent and one also has the bounds for the parameters
$\lambda_{2}$ and $\alpha_{2}$,

\begin{equation}
 {\alpha_{2} \over \alpha_{1}^2} \leq 10^6 \,\,\,,
{\lambda_{2} \over \alpha_{1}^2} \leq 10^3 \nonumber \,,
\end{equation}

\noindent which gives a restriction on the aditional mass 
scale $M_{E}$, ${M_{E}^2 \over M_{Pl}^2} \geq 10^{-3}$ . 
 
\section{Summary}

It has been shown that the perturbative renormalization of the 
theory of a scalar field coupled to the gravitational field, 
which requires an infinite number of counterterms, is compatible 
with the presence of only a finite number of renormalized parameters 
which can be choosen arbitrarily at a given scale. In this sense 
one has a theory with the same predictibility as a renormalizable 
theory in the power counting sense. The corrections to the low 
energy limit due to the higher order terms in the energy expansion 
of the effective theory do not necessarily involve additional free
parameters. This result found in the case of pure 
gravity~\cite{AC} has been extended to the case of a 
scalar field coupled to gravity in two different forms (minimal and 
non-minimal reduction of couplings). The effective theory has several 
characteristic mass scales: the scale associated to the classical 
limit (the Planck mass $M_{Pl}$ ), the scale associated to the 
fluctuations of the geometry ( $M_{R}$ ), the scale of the 
fluctuations of the matter system ( $M_{E}$ ), the scale limiting 
the validity of the perturbative treatment ( $M_{0}$ ) and the mass 
of the scalar field ( $m$ ). Each of these scales is in one to one 
correspondence with the independent dimensionless parameters of the 
effective theory after a reduction of couplings compatible with the 
renormalization group equations. The reduction of couplings puts some 
limitations on the mass scales. The reduction of the couplings in the 
gravitational lagrangian requires  
${\alpha_{2} \over \alpha_{1}^2} \gg 1$
and then $M_{R} \ll M_{Pl}$ . The treatment of the scalar mass as
a perturbation requires $\lambda_{-2} \lambda_{2} \ll 1$ which
implies that $m \ll M_{E}$ . In the case of minimal reduction 
one has $\lambda_{2} \sim \alpha_{1}^{2}$ and then $M_{E} \sim M_{Pl}$ .

In order to have an energy range within the perturbative domain where
some of the low energy limit corrections have to be included, it is
necessary for either $M_{R}$ or $M_{E}$ to be smaller than $M_{0}$.
In the case of minimal reduction of couplings the contribution from 
a term ${\cal L}_m^{(n)}$ in the matter lagrangian, which is of order
${\lambda_{2n} \over M^{2n}} E^{2n}$ when one is considering the matter 
system at energies of order $E$, gives an effect of order
$\left({E \over M_{Pl}}\right) \left({E^2 \over M_{R}^2} \right)^{n-1}$ .
Then even if the scale $M_{0}$ for the domain of validity of the 
perturbative treatment can be much lower than $M_{Pl}$ one has to
consider energies $E \sim M_{R}$ in order to go beyond the low energy
limit.

It is natural to expect that the results found for a scalar field 
coupled to gravity can be generalized to more general cases for the
matter system. Instead of the scalar self-coupling $\lambda_{0}$
one will have a set of dimensionless couplings for the non 
gravitational interactions and instead of the mass of the scalar one 
will have one or several masses for the matter fields.

The cosmological constant problem has been reformulated at the level
of the renormalization group equations. In the presence of a mass
for the scalar field the (logarithmic) derivative of the cosmological
parameter with respect to the renormalization scale has a contribution
which is not proportional to the cosmological parameter. Then it is
not possible to have an arbitrarily small value for this parameter
over a given range of scales. Two possible solutions of this problem
have been considered corresponding to the possibility that these
terms in the $\beta$-function of the cosmologic parameter which are
not proportional to it either vanish for an appropriate choice of matter
lagrangian or are sufficiently small to be compatible with the 
smallness of the cosmological constant. The number of conditions that
the matter system have to satisfy, in order to have a solution
to the cosmological constant problem based on the second possibility
mentioned above, is significantly reduced if one considers an
effective theory with a reduction of couplings.

\bigskip

\bigskip\leftline{{\bf Acknowledgments}}\medskip
This work was partially supported by the CICYT (proyecto AEN 94--0218).  
The work of M.A. has been supported by a DGA fellowship.
 
\vfill
\eject

\newpage

\end{document}